%% file: main.tex
\documentclass[a4paper]{article}

\usepackage{INTERSPEECH2022}
\usepackage{amsmath,graphicx}
\usepackage{amssymb}
\usepackage{bbm}
\usepackage{multirow}
\usepackage{pifont}
\usepackage{subcaption}

\title{Semi-supervised Time Domain Target Speaker Extraction with Attention}
\name{Zhepei Wang$^{\sharp *}$ \thanks{* Part of the work performed while at Amazon Web Services.}, Ritwik Giri$^{\dagger}$, Shrikant Venkataramani$^{\dagger}$, Umut Isik$^{\dagger}$, Jean-Marc Valin$^{\dagger}$, Paris Smaragdis$^{\sharp \dagger}$, Mike Goodwin$^{\dagger}$, Arvindh Krishnaswamy$^{\dagger}$}
\address{
  $^\sharp$University of Illinois at Urbana-Champaign\\
  $^\dagger$Amazon Web Services}
\email{zhepeiw2@illinois.edu, ritwikg@amazon.com}

\begin{document}
\input{commands}

\maketitle
\begin{abstract}
In this work, we propose Exformer, a time-domain architecture for target speaker extraction. It consists of a pre-trained speaker embedder network and a separator network based on transformer encoder blocks. We study multiple methods to combine speaker information with the input mixture, and the resulting Exformer architecture obtains superior extraction performance compared to prior time-domain networks. Furthermore, we investigate a two-stage procedure to train the model using mixtures without reference signals upon a pre-trained supervised model. Experimental results show that the proposed semi-supervised learning procedure improves the performance of the supervised baselines.

\end{abstract}

\input{intro}
\input{model}
\input{semisup}
\input{experiments}
\input{results}

\input{conclusion}

\bibliographystyle{IEEEtran}
\newpage
\bibliography{mybib}

\end{document}

%% file: commands.tex
\newcommand{\R}{\mathbb{R}}
\newcommand{\bD}{\mathbf{D}}
\newcommand{\bH}{\mathbf{H}}
\newcommand{\bM}{\mathbf{M}}
\newcommand{\bS}{\mathbf{S}}
\newcommand{\bU}{\mathbf{U}}
\newcommand{\bV}{\mathbf{V}}
\newcommand{\bW}{\mathbf{W}}
\newcommand{\bZ}{\mathbf{Z}}
\newcommand{\ba}{\mathbf{a}}
\newcommand{\bh}{\mathbf{h}}
\newcommand{\bq}{\mathbf{q}}
\newcommand{\bs}{\mathbf{s}}
\newcommand{\bv}{\mathbf{v}}
\newcommand{\bx}{\mathbf{x}}
\newcommand{\by}{\mathbf{y}}
\newcommand{\bz}{\mathbf{z}}
\newcommand{\cmark}{\ding{51}}%
\newcommand{\xmark}{\ding{55}}%

%% file: intro.tex
\section{Introduction}
\label{sec:intro}

Communications in the real world often take place in complex auditory scenes, where the speech of a target speaker is likely to be contaminated by an interfering speaker or background noise. To improve the quality of speech, deep learning methods have been developed for target speaker extraction (TSE) with promising performance \cite{VF, speakerbeam}. As opposed to blind source separation (BSS) systems where all individual sources are separated by the model, TSE systems isolate the speech signal of the speaker in interest from all other speakers and ambiance noise. By focusing only on the target speaker and treating the rest of the signal as interference, TSE systems do not require the information of the total number of sources beforehand and do not suffer from the permutation ambiguity, which are the two primary challenges in BSS \cite{deepclustering, pit}.

The VoiceFilter \cite{VF} is one of the first neural-based architectures to extract the speech of a target speaker. It contains a bidirectional LSTM (BLSTM)-based separation module to extract the magnitude mask of the target signal from the mixture spectrogram. A pre-trained speaker embedding module extracts cues of the target speaker, which requires an enrollment utterance of the target speaker as an additional input. The speaker embedding guides the separator during the mask estimation. Several studies follow this embedder-separator design with improvements in the architecture to enhance the extraction quality or running-time efficiency using spectral \cite{VF_lite, speakerbeam, sbf_mtsal} or perceptual \cite{personalized_percepnet} features as input.

The recent success of end-to-end networks such as ConvTasNet \cite{convtasnet} in BSS has drawn attention to time-domain architectures in TSE. The separator includes paired 1D convolutional encoder and decoder, along with stacks of convolutional blocks in the middle. The embedder network can be pre-trained \cite{TencentTSE} or learned jointly with the separator \cite{td_speakerbeam, speakerfilter, time_tse_mult, spex, spex+, spex++}. Inspired by the Sepformer \cite{sepformer} in BSS, we propose the Exformer and study the effectiveness of the multi-head attention (MHA) mechanisms with the transformer encoder \cite{transformer} as the building block of the separator module for time-domain TSE. Specifically, we compare multiple configurations of the fusion of the speaker embedding into the separator module within the transformer blocks.

Despite the performance gain from the architectural improvement, these networks usually require a vast amount of labeled data to train. While it is unrealistic to obtain both the mixture and its source components in real-world acoustic conditions, noisy speech recordings exist on a large scale and are easy to collect. In the mixture invariant training (MixIT) \cite{mixit}, separation models are trained with mixture of mixtures in both unsupervised and semi-supervised setups. An alternative approach to leverage the noisy mixtures is to train the network with pseudo-labels assigned by a pre-trained teacher model, with promising results in singing-voice separation \cite{semisup_svs, improved_svs} and speech enhancement \cite{remixit, personalized_se}. Meanwhile, training with noisy data in TSE is yet to be explored. We propose a semi-supervised framework to train the Exformer with both labeled and unlabeled data. Empirical studies on two-speaker mixtures show that the proposed Exformer outperforms the prior Conv-TasNet-based architectures under supervised learning, and the incorporation of unlabeled data with the semi-supervised setup further improves the quality of the extracted signal.

%% file: model.tex
\section{Target Speaker Extraction Model}
\label{sec:model}
\begin{figure*}[ht]
  \centering
  \includegraphics[width=0.8\linewidth]{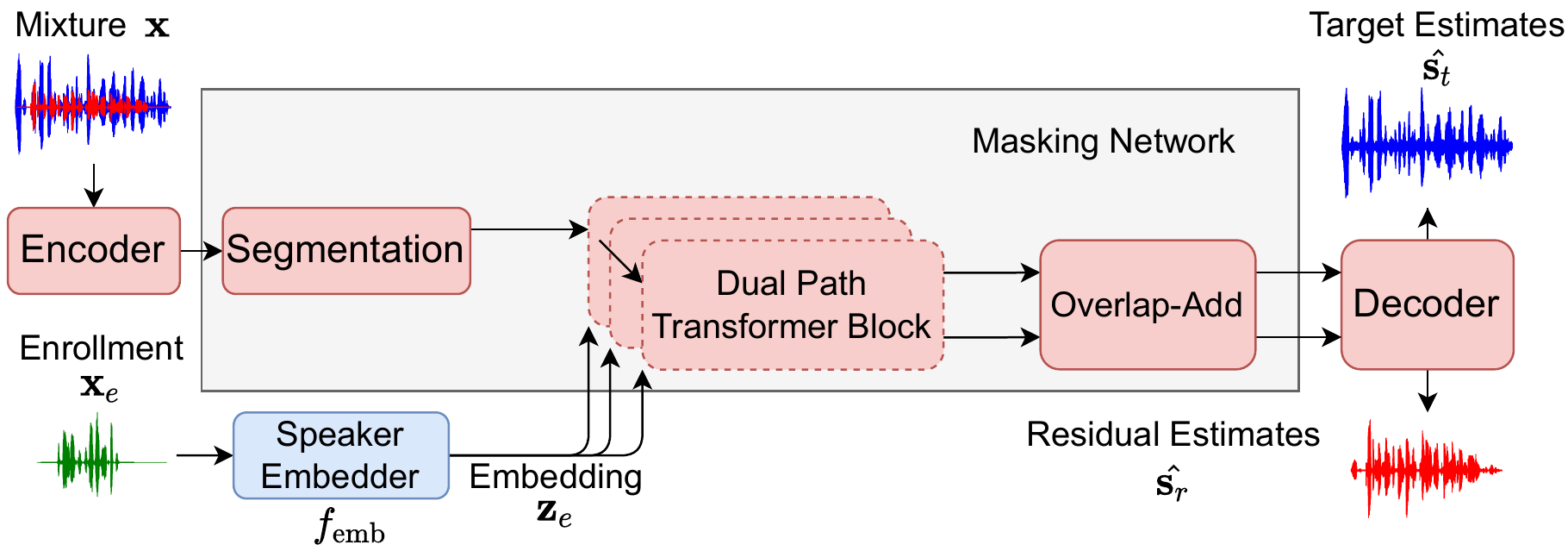}
\caption{The architecture of the Exformer. The speaker embedder is pre-trained, and its weights are fixed during the training of the extraction network. Speaker embeddings are applied to each of the dual-path transformer blocks.}
\label{fig:exformer}
\end{figure*}

The proposed Exformer model (Figure~\ref{fig:exformer}) is based upon the time domain speech separation network Sepformer \cite{sepformer} with a speech encoder, masking network, and a speech decoder. A pre-trained BLSTM speaker embedder is used to extract speaker information to guide the extraction. The speaker extractor (encoder, masking network, and decoder) is trained with the negative scale-invariant signal-to-distortion ratio (SI-SDR) \cite{sisdr}, and the parameters of the embedder network are fixed.

\subsection{Speaker Embedder}
\label{ssec:model_emb}
The embedder network obtains a fix-sized embedding vector $\bz_e = f_\text{emb}(\bx_e) \in \R^D$ of the target speaker, where $f_\text{emb}$ is the embedder network and $\bx_e$ is the enrollment utterance. Following the embedder design in \cite{VF}, the mel-spectrogram of the enrollment speech is processed by a stack of BLSTM layers to model sequential information. The average of the hidden-state representation from the first and the last frames is projected to a lower dimension with a linear layer. Finally, the projection output is normalized with unit $l2$ norm to obtain the speaker embedding $\bz_e$. The embedder network is pre-trained with the generalized end-to-end loss (GE2E) \cite{ge2e}.

\subsection{Encoder}
\label{ssec:model_enc}

The encoder network consists of a 1D convolutional layer followed by ReLU activation. It transforms the time domain input mixture $\bx \in \R^L$ with $L$ samples to a hidden representation $\bH \in \R^{F \times T}$ with $F$ features and $T$ temporal frames.



\subsection{Masking Network}
\label{ssec:model_mask}

The masking network takes the mixture representation $\bH$ and the speaker embedding $\bz_e$ as input and estimates a mask $\bM_c \in \R^{F \times T}, c \in \{1, 2, \ldots C\}$, where $C$ is the total number of sources. We set $C=2$ as the model estimates one mask for the target speech and another mask for the residual interfering signal. Similar to DPRNN \cite{dprnn} and Sepformer \cite{sepformer}, the masking network consists of three modules: segmentation, embedding-guided dual-path processing, and overlap-add.

\subsubsection{Segmentation}
\label{sssec:model_mask_seg}
The segmentation module divides the $T$ frames from the mixture representation $\bH$ into $S$ segments, each with length $K$, with 50\% overlap between segments. All segments are concatenated to form a tensor $\bV^{(0)} = \{\bV^{(0)}_s\}_{s=1}^S \in \R^{F\times K \times S}$.

\subsubsection{Embedding Fusion}
\label{sssec:model_mask_emb}
The embedding fusion module combines the segmented mixture representation and the target speaker embedding $\bz_e$. We apply embedding fusion at the beginning of each processing block as in \cite{spex, spex+, speakerfilter, time_tse_mult}.  At the beginning of block $j$, we perform
\begin{align}
    \bW^{(j)} = f_\text{fus}^{(j)}(\bV^{(j-1)}, \bz_e),
\end{align}
where $\bW^{(j)} \in \R^{F\times T \times S}$ is the embedding-guided mixture representation,  $f_\text{fus}^{(j)}$ is the fusion function that applies independently to each segment and frame, and the superscript $j$ denotes the block index. 

We experiment with multiple approaches to incorporate speaker information into the separation pipeline. The first method concatenates the embedding and the latent mixture representation similar to \cite{VF, tsenet, spex, spex+, spex++, speakerfilter}. The embedding $\bz_e$ is first duplicated on the frame and the chunk dimensions to obtain $\bZ_e \in \R^{D\times K \times S}$. Then, $\bZ_e$ is concatenated along the feature dimension to $\bV^{(j-1)}$ and linearly projected back to the $F$-dimensional space:
\begin{align}
    f_\text{cat}^{(j)}(\bV^{(j-1)}, \bZ_e) = \text{Linear}(\text{Concat}(\bV^{(j-1)}, \bZ_e)).
    \label{eq:cat}
\end{align}

In the second approach, we first project the embedding $\bZ_e$ to the same dimension as $\bV^{(j-1)}$ and then element-wise multiply with the mixture representation following \cite{time_tse_mult}:
\begin{align}
    f_\text{mult}^{(j)}(\bV^{(j-1)}, \bZ_e) = \text{Linear}(\bZ_e) \odot \bV^{(j-1)},
    \label{eq:mult}
\end{align}
where $\odot$ stands for element-wise multiplication.

We also consider replacing the element-wise multiplication with addition similar to the application of positional embeddings in transformer \cite{transformer}, leading to our third approach:
\begin{align}
    f_\text{add}^{(j)}(\bV^{(j-1)}, \bZ_e) = \text{Linear}(\bZ_e) + \bV^{(j-1)}.
    \label{eq:add}
\end{align}

\subsubsection{Dual-Path Processing}
\label{sssec:model_mask_dualpath}

\begin{figure}
  \begin{subfigure}[t]{0.375\columnwidth}
    \includegraphics[width=\linewidth]{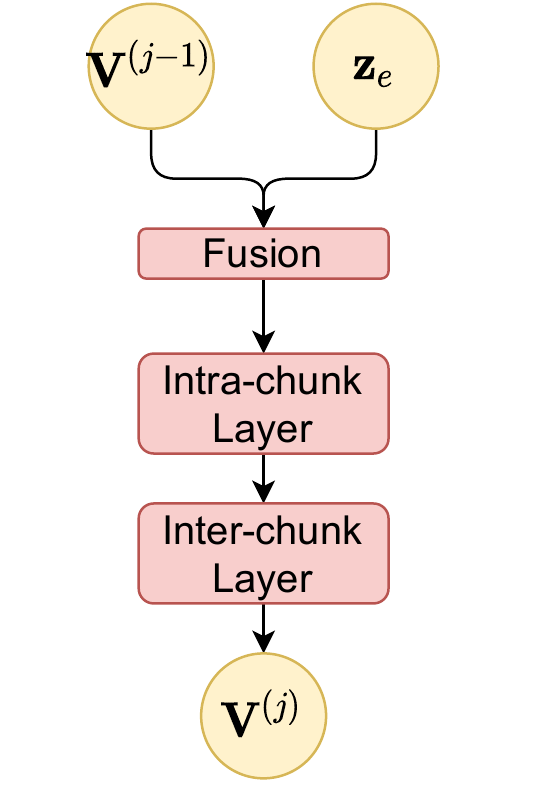}
    \caption{The $j$-th dual path transformer block.}
    \label{fig:dp}
  \end{subfigure}
  \hfill 
  \begin{subfigure}[t]{0.625\columnwidth}
    \includegraphics[width=\linewidth]{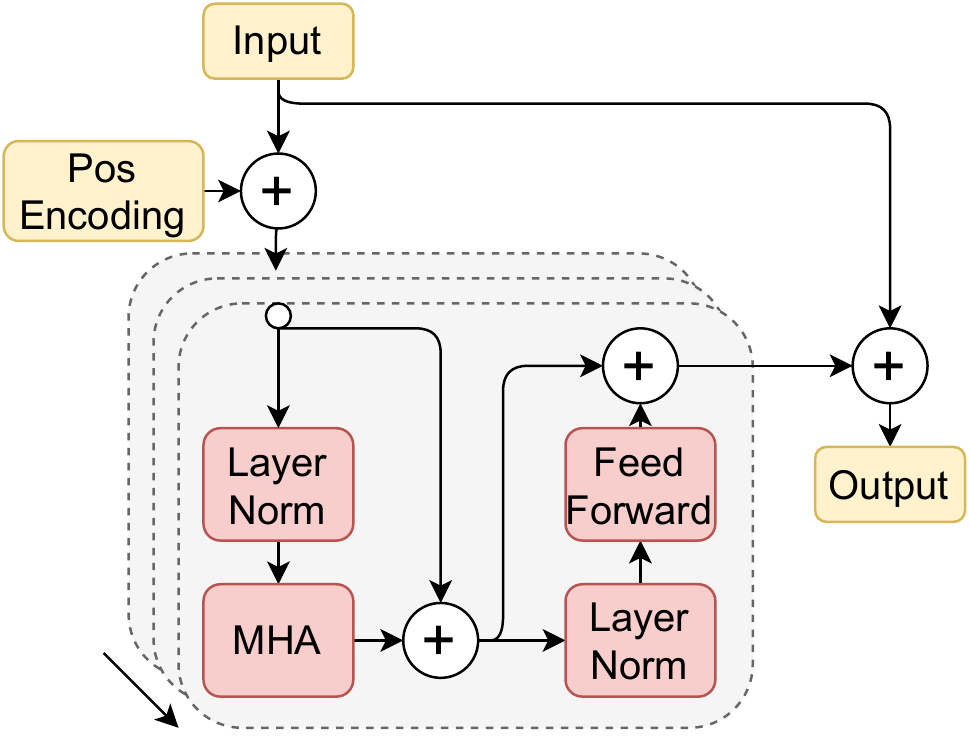}
    \caption{An intra (inter)-chunk layer.}
    \label{fig:intra}
  \end{subfigure}
  \caption{The dual-path processing module is made up of a stack of dual-path transformer blocks, where each block contains a fusion layer for processing speaker information, an intra-chunk layer for learning local dependency, and an inter-chunk layer for modeling global dependency.}
\end{figure}

Figure ~\ref{fig:dp} shows the pipeline of a dual-path transformer block. After the embedding fusion, the dual-path module alternates the processing between short-term and long-term information. The intra-chunk layer computes the local dependency within each segment, while the inter-chunk layer communicates between the segments to model global information. As shown in Figure~\ref{fig:intra}, each of the intra-chunk and inter-chunk layers contains multiple transformer encoder blocks \cite{transformer}, including layer normalization \cite{layernorm}, multi-head attention (MHA), and a feed-forward layer with residual connections. A sinusoidal positional encoding is applied before each MHA layer. Each transformer block preserves the number of features $F$.

The intra-chunk layer operates on each of the $S$ segments independently and computes the local dependency within each segment. For the $j$-th dual-path block, the intra-chunk layer is applied to the second dimension of the speaker-guided input tensor $\bW^{(j)}$ by 
\begin{align}
    \bU^{(j)}_i = f_\text{intra}^{(j)}(\bW^{(j)}[:, :, i]), i \in \{1, 2, \ldots, K\},
    \label{eq:intra}
\end{align}
where $f_\text{intra}^{(j)}$ represents the $j$-th intra-chunk layer, and the intra-chunk output $\bU^{(j)}$ is obtained by concatenating $\{\bU_i^{(j)}\}_{i=1}^K$ along the third dimension. The inter-chunk layer processes the long-term, global dependency between the segments and is applied to the third dimension of the intra-chunk layer's output
\begin{align}
    \bV^{(j)}_i = f_\text{inter}^{(j)}(\bU^{(j)}[:, i, :]), i \in \{1, 2, \ldots, S\}, 
    \label{eq:inter}
\end{align}
and we concatenate $\{\bV_i^{(j)}\}_{i=1}^S$ along the second dimension to obtain the block output $\bV^{(j)}$, which is further processed by block $j+1$. With the interleaving intra-chunk and inter-chunk layers, the model effectively models both local and global dependency in turn. 

With a total of $B$ dual-path blocks, the output $\bV^{(B)}$ is processed by a 2D convolutional layer to increase its dimension by $C$ folds, representing each of the $C$ output sources with $\bM' = \{\bM'_c\}_{c=1}^C = \text{Conv2D}(\bV^{(B)})$, where $\bM'_c \in \R^{F \times K \times S}$.

\subsubsection{Overlap-add}
\label{sssec:model_mask_overlap}
Finally, an overlap-add operation transforms the $S$ segments in each $\bM'_c$ back to $T$ frames followed by a ReLU activation to obtain the estimated mask $\bM_c \in \R^{F \times T}$ for source $c$.

\subsection{Decoder}
\label{ssec:model_dec}

The decoder is a 1D transposed convolutional layer that transforms the masked representation $\bH_c = \bM_c \odot \bH \in \R^{F \times T}$ of each source $c$ to the estimation of the time domain signal $\hat{\bs_c}= \text{TransposedConv1D}(\bH_c) \in \R^L$.




%% file: semisup.tex
\section{Semi-supervised Setup}
\label{sec:semisup}
Prior works on TSE focus on supervised training with the reference speech available. To incorporate a large unlabeled corpus, where the ground-truth target is missing, we propose a semi-supervised approach by applying a triplet speaker representation loss \cite{srl}. Given the estimated target signal $\hat{\bs_t}$ and residual $\hat{\bs_r}$, we compute the embeddings $\bz_t = f_\text{emb}(\hat{\bs_t})$ and $\bz_r = f_\text{emb}(\hat{\bs_r})$ with the pre-trained embedder, and compute their distances to $\bz_e$ which is obtained from the enrollment speech. If the estimated target is separated from the residual, $\bz_e$ should be much more similar to $\bz_t$ than to $\bz_r$. The triplet embedder loss is defined as
\begin{align}
    \mathcal{L}_\text{emb}(\bz_e, \bz_t, \bz_r) = \max \{\lVert \bz_e - \bz_t \lVert_2 - \lVert \bz_e - \bz_r \rVert_2 + \gamma, 0\},
    \label{eq:emb_loss}
\end{align}
where $\gamma=1$ represents the margin. The embedder loss is computed along with the reconstruction loss in a supervised context in \cite{srl}, while we extend its application to unlabeled data. Specifically, when the ground-truth signal is not available, we only compute the embedder loss; otherwise, we compute both the reconstruction loss $\mathcal{L}_\text{SI-SDR}$ and $\mathcal{L}_\text{emb}$. The overall objective for the semi-supervised setup is
\begin{align}
\begin{split}
    \mathcal{L}_\text{semi} & = \lambda_s \mathbbm{1}[\{\bs, \bx, \bx_e\}\in \mathcal{D}_s] \cdot  \frac{1}{2}\sum_{c \in \{t, r\}} \mathcal{L}_\text{SI-SDR}(\bs_c, \hat{\bs_c})) \\
    & + \lambda_u \mathcal{L}_\text{emb}(\bz_e, \bz_t, \bz_r),
\end{split}
\label{eq:semi}
\end{align}
where $\lambda_s, \lambda_u$ are hyperparameters controlling the two losses, $\mathbbm{1}$ is the indicator function, and $\mathcal{D}_s$ represents the dataset with ground-truth target speech.

One concern with applying semi-supervised training from scratch is that it might be difficult for the model to obtain optimal separation performance quickly since the parameters are drifted away by the embedder loss (\ref{eq:emb_loss}). We experiment with a two-stage setup where the model is first trained with SI-SDR using only labeled data and then trained with both labeled and unlabeled data with (\ref{eq:semi}). In the supervised stage, the model learns to adapt the speaker embedding and to extract the target speech for high-quality reconstruction. When the performance reaches the plateau, we enter the second stage by introducing the unlabeled data and the embedder loss. The goal for the second stage is to improve the model's ability to generalize with the additional unlabeled data.

%% file: experiments.tex
\section{Experimental Setup}
\label{sec:exp}

\subsection{Dataset}
\label{ssec:exp_dataset}
We train and evaluate the system with the simulated 2-speaker mixture from the \textit{WSJ0-2mix-extr}\footnote{\text{https://github.com/xuchenglin28/speaker\_extraction}} dataset. The recordings are at a sampling rate of 8 kHz. The training, development, and test set contains 20,000, 5,000, and 3,000 utterances, respectively. The mixtures are generated from speech recordings at a signal-to-noise ratio (SNR) between 0 and 5 dB. The first speaker is chosen to be the target speaker and the second speaker is treated as interference. For each target speaker, another speech recording from this speaker is randomly selected as the enrollment utterance. For the semi-supervised setup, we incorporate over 100,000 recordings from the VoxCeleb 1 corpus \cite{voxceleb}. Since the recordings may contain background and reverberation, they are not suitable for the standard supervised setup.

\subsection{Model and Training Details}
\label{ssec:exp_train}
First, we pre-train the speaker embedder following \cite{VF}. The embedder takes the 40-dimensional log mel-spectrogram as input. It consists of 3 BLSTM layers, each with 768 hidden units, and a linear projection head is attached to obtain an embedding with size 256. The network is trained on the VoxCeleb 1 \& 2 \cite{voxceleb} and the clean training partition of the Librispeech \cite{librispeech} corpus combined, which may contain ambiance noise or reverberation. To adapt to the memory constraint, each batch consists of 46 speakers, each speaker with 4 utterances of 4 seconds long. The network parameters are updated with the Adam optimizer \cite{adam} with an initial learning rate of 5e-4. During preliminary experiments on supervised TSE, we obtain the best performing teacher network with the embedder checkpoint at 550,000 steps with an equal error rate (EER) of 6.5\% on the VoxCeleb1 verification test set.

We train the extractor network with the optimal configuration from \cite{sepformer}. The encoder convolution has a kernel size of 16 and a stride of 8 with 256 output channels. The masking network contains two dual-path blocks. Each intra-chunk or inter-chunk module contains 8 transformer encoder layers, each layer with 8-head attention and a 1024-dimensional feed-forward projection. The decoder mirrors the encoder with the same kernel and stride sizes. We experiment with concatenation, element-wise multiplication and addition as the embedding-mixture fusion discussed in \ref{sssec:model_mask_emb}. 

During training, we apply dynamic mixing \cite{sepformer} to generate mixtures on the fly from a random 3-second segment from each source recording. If either of the sources is less than three seconds, zero padding is applied on both ends. We also apply time-domain speed perturbing by 5\% as data augmentation. We set the batch size as 1 for the experiments, and we use an Adam optimizer with an initial learning rate of 1.5e-4. The learning rate is reduced by half if the validation loss is not improved for two consecutive epochs after 85 epochs until the learning rate reaches 1e-6.

For the semi-supervised training setup, we weight the two loss functions with $\lambda_s=1$ and $\lambda_u=0.05$. We experiment with different probabilities of data sampled from the labeled versus unlabeled corpus. For the two-stage setup, we proceed with the best-performing model from the supervised setup. We adjust the initial learning rate for the second stage to be 7.5e-5, and the aforementioned learning rate scheduler applies after 65 epochs. We encourage the readers to refer to our implementation\footnote{\text{https://github.com/zhepeiw/tse\_semi}}.


%% file: results.tex
\section{Results and Discussions}
\label{sec:res}

We compare the different configurations of the proposed Exformer with previous works using the test partition of the WSJ0-2mix-extr dataset. We evaluate each system with the SI-SDR and the perceptual evaluation of speech quality (PESQ) \cite{pesq} score of the model's output with respect to the reference speech.

\subsection{Supervised Training and Fusion Configuration}
\label{ssec:res_sup}


\begin{table}[ht]
    \caption{Performance on the WSJ0-2mix-extr test set of the proposed and baseline time-domain TSE systems under supervised training, with the best performance in bold.}
    \centering
    \begin{tabular}{c|c|c|c}
        Method & Fusion Type & SI-SDR (dB) & PESQ \\
        \hline
         Input & \xmark & 2.51 & 1.81 \\
        \hline
        TseNet \cite{tsenet} & \multirow{2}{*}{Concat} & 14.73 & 3.14 \\
        SpEx+ \cite{spex+} &  & 18.2 & 3.49 \\
        \hline
        \multirow{3}{*}{Exformer (Ours)} & Concat & 19.05 & 3.78 \\
        & Mult & 19.55 & \textbf{3.82} \\
        & Add & \textbf{19.85} & \textbf{3.82} \\
        \hline
    \end{tabular}
    
    \label{tab:sup}
\end{table}

Table~\ref{tab:sup} summarizes the performance of previous methods on time-domain target speech extraction as well as the proposed architecture with multiple configurations under the supervised setup. We report the mean SI-SDR and PESQ for each setup. The statistics for the baselines are adapted from \cite{spex+}.

All three variants of the Exformer outperform the baseline methods \cite{tsenet, spex+}. The performance gain can be attributed to the use of the dual-path transformer blocks in the masking network which effectively model local and global dependency. Our observations are consistent with the findings of BSS systems in \cite{sepformer}, where the performance of the transformer blocks dominates over that of the convolutional blocks.

We also study the impact of different fusion strategies for speaker embedding and mixture representation. Among the three fusion types, additive fusion has the best performance with a gain of 0.3 dB SI-SDR over element-wise multiplication and a 0.8dB gain over concatenation. Addition and multiplication share the same PESQ score slightly higher than concatenation. 
We show in our experiments the effectiveness of adding a time-invariant speaker embedding to the latent representation, allowing the extraction network to effectively capture the conditioning speaker information.

\subsection{Semi-supervised Training and Embedder Loss}
\label{ssec:res_semi}

\begin{table}[ht]
    \caption{Results of the Exformer trained with the embedder loss along with multiple semi-supervised configurations.}
    \centering
    \begin{tabular}{c|c|c|c|c}
    Embedder & Unlabeled & \multirow{2}{*}{Two-stage} & SI-SDR & \multirow{2}{*}{PESQ} \\
    Loss & Percentage & & (dB) & \\
    \hline
      \xmark   & \multirow{3}{*}{0} & \xmark & 19.85 & 3.82 \\
      \cmark   &  & \xmark & 19.74 & \textbf{3.85} \\
      \cmark & & \cmark & 20.21 & 3.83 \\
      \hline
      \multirow{2}{*}{\cmark}   & \multirow{2}{*}{5} & \xmark & 19.52 & 3.79 \\
       &  & \cmark & 20.34 & 3.83 \\
       \hline
       \multirow{2}{*}{\cmark}   & \multirow{2}{*}{10} & \xmark & 19.17 & 3.74 \\
       &  & \cmark & \textbf{20.49} & \textbf{3.85} \\
       \hline
      \multirow{2}{*}{\cmark}   & \multirow{2}{*}{25} & \xmark & 19.60 & 3.82 \\
       &  & \cmark & 20.20 & 3.83 \\
       \hline
       \multirow{2}{*}{\cmark}   & \multirow{2}{*}{50} & \xmark & 17.27 & 3.56 \\
       &  & \cmark & 19.92 & 3.82  \\
       \hline
    \end{tabular}
    
    \label{tab:semi}
\end{table}

We next study the impact of the embedder loss (\ref{eq:emb_loss}) and the use of unlabeled data. We report in Table~\ref{tab:semi} the performance of the Exformer trained on different combinations of the loss functions, number of training stages, and the probability of unlabeled data being sampled into a training batch. We initialize the networks in the second stage with the weights of the best performing supervised system in Table~\ref{tab:sup}.

Training the network from scratch using the embedder loss yields worse performance. This observation verifies our hypothesis that the embedder loss does not contribute to optimal separation performance with random initialization compared to training with reconstruction loss alone. With two-stage training, however, the extraction performance improves consistently among various percentages of applying unlabeled data. As the percentage increases, the performance reaches the peak at a probability of 10\% of sampling unlabeled data in both evaluation metrics with a 0.6 dB gain of SI-SDR over the supervised baseline. The results show that the system benefits from a two-stage procedure with the embedder loss using a small ratio of unlabeled to labeled data.

%% file: conclusion.tex
\section{Conclusion}
\label{sec:conclusion}
We propose the Exformer, a time-domain transformer-based architecture for target speaker extraction. Under the supervised training setup, the Exformer significantly outperforms prior time-domain networks. We further show that the extraction performance can be enhanced with a two-stage semi-supervised pipeline incorporating mixtures without reference signals. In future work, we would explore target speaker extraction under noisy and reverberant conditions and investigate approaches to include massive unlabeled data such as curriculum learning.

\textbf{Acknowledgement:} This work is partially funded by NIFA award \#2020-67021-32799.

%% file: main.bbl
\begin{thebibliography}{10}
\providecommand{\url}[1]{#1}
\csname url@samestyle\endcsname
\providecommand{\newblock}{\relax}
\providecommand{\bibinfo}[2]{#2}
\providecommand{\BIBentrySTDinterwordspacing}{\spaceskip=0pt\relax}
\providecommand{\BIBentryALTinterwordstretchfactor}{4}
\providecommand{\BIBentryALTinterwordspacing}{\spaceskip=\fontdimen2\font plus
\BIBentryALTinterwordstretchfactor\fontdimen3\font minus
  \fontdimen4\font\relax}
\providecommand{\BIBforeignlanguage}[2]{{%
\expandafter\ifx\csname l@#1\endcsname\relax
\typeout{** WARNING: IEEEtran.bst: No hyphenation pattern has been}%
\typeout{** loaded for the language `#1'. Using the pattern for}%
\typeout{** the default language instead.}%
\else
\language=\csname l@#1\endcsname
\fi
#2}}
\providecommand{\BIBdecl}{\relax}
\BIBdecl

\bibitem{VF}
Q.~Wang, H.~Muckenhirn, K.~Wilson, P.~Sridhar, Z.~Wu, J.~R. Hershey, R.~A.
  Saurous, R.~J. Weiss, Y.~Jia, and I.~L. Moreno, ``{VoiceFilter: Targeted
  Voice Separation by Speaker-Conditioned Spectrogram Masking},'' in
  \emph{Interspeech}, 2019.

\bibitem{speakerbeam}
K.~Žmolíková, M.~Delcroix, K.~Kinoshita, T.~Ochiai, T.~Nakatani, L.~Burget,
  and J.~Černocký, ``Speakerbeam: Speaker aware neural network for target
  speaker extraction in speech mixtures,'' \emph{IEEE Journal of Selected
  Topics in Signal Processing}, vol.~13, no.~4, 2019.

\bibitem{deepclustering}
J.~R. Hershey, Z.~Chen, J.~L. Roux, and S.~Watanabe, ``Deep clustering:
  Discriminative embeddings for segmentation and separation,'' \emph{IEEE
  International Conference on Acoustics, Speech and Signal Processing
  (ICASSP)}, 2016.

\bibitem{pit}
D.~Yu, M.~Kolb{\ae}k, Z.~Tan, and J.~H. Jensen, ``Permutation invariant
  training of deep models for speaker-independent multi-talker speech
  separation,'' \emph{IEEE International Conference on Acoustics, Speech and
  Signal Processing (ICASSP)}, 2017.

\bibitem{VF_lite}
Q.~Wang, I.~L. Moreno, M.~Saglam, K.~Wilson, A.~Chiao, R.~Liu, Y.~He, W.~Li,
  J.~Pelecanos, M.~Nika, and A.~Gruenstein, ``{VoiceFilter-Lite: Streaming
  Targeted Voice Separation for On-Device Speech Recognition},'' in
  \emph{Interspeech}, 2020.

\bibitem{sbf_mtsal}
C.~Xu, W.~Rao, C.~E. Siong, and H.~Li, ``Optimization of speaker extraction
  neural network with magnitude and temporal spectrum approximation loss,''
  \emph{IEEE International Conference on Acoustics, Speech and Signal
  Processing (ICASSP)}, 2019.

\bibitem{personalized_percepnet}
R.~Giri, S.~Venkataramani, J.-M. Valin, U.~Isik, and A.~Krishnaswamy,
  ``Personalized percepnet: Real-time, low-complexity target voice separation
  and enhancement,'' \emph{Interspeech}, 2021.

\bibitem{convtasnet}
Y.~Luo and N.~Mesgarani, ``Conv-tasnet: Surpassing ideal time–frequency
  magnitude masking for speech separation,'' \emph{IEEE/ACM Transactions on
  Audio, Speech, and Language Processing (TASLP)}, vol.~27, 2019.

\bibitem{TencentTSE}
X.~Ji, M.~Yu, C.~Zhang, D.~Su, T.~Yu, X.~Liu, and D.~Yu, ``Speaker-aware target
  speaker enhancement by jointly learning with speaker embedding extraction,''
  in \emph{IEEE International Conference on Acoustics, Speech and Signal
  Processing (ICASSP)}, 2020.

\bibitem{td_speakerbeam}
M.~Delcroix, T.~Ochiai, K.~Žmol{\'i}kov{\'a}, K.~Kinoshita, N.~Tawara,
  T.~Nakatani, and S.~Araki, ``Improving speaker discrimination of target
  speech extraction with time-domain speakerbeam,'' \emph{IEEE International
  Conference on Acoustics, Speech and Signal Processing (ICASSP)}, 2020.

\bibitem{speakerfilter}
S.~He, H.~Li, and X.~Zhang, ``Speakerfilter: Deep learning-based target speaker
  extraction using anchor speech,'' in \emph{IEEE International Conference on
  Acoustics, Speech and Signal Processing (ICASSP)}, 2020.

\bibitem{time_tse_mult}
J.~Zhao, S.~Gao, and T.~Shinozaki, ``{Time-Domain Target-Speaker Speech
  Separation with Waveform-Based Speaker Embedding},'' in \emph{Interspeech},
  2020.

\bibitem{spex}
C.~Xu, W.~Rao, E.~S. Chng, and H.~Li, ``Spex: Multi-scale time domain speaker
  extraction network,'' \emph{IEEE/ACM Transactions on Audio, Speech, and
  Language Processing (TASLP)}, vol.~28, 2020.

\bibitem{spex+}
M.~Ge, C.~Xu, L.~Wang, C.~E. Siong, J.~Dang, and H.~Li, ``{SpEx+: A Complete
  Time Domain Speaker Extraction Network},'' in \emph{Interspeech}, 2020.

\bibitem{spex++}
------, ``Multi-stage speaker extraction with utterance and frame-level
  reference signals,'' \emph{IEEE International Conference on Acoustics, Speech
  and Signal Processing (ICASSP)}, 2021.

\bibitem{sepformer}
C.~Subakan, M.~Ravanelli, S.~Cornell, M.~Bronzi, and J.~Zhong, ``Attention is
  all you need in speech separation,'' \emph{IEEE International Conference on
  Acoustics, Speech and Signal Processing (ICASSP)}, 2021.

\bibitem{transformer}
A.~Vaswani, N.~Shazeer, N.~Parmar, J.~Uszkoreit, L.~Jones, A.~N. Gomez, L.~u.
  Kaiser, and I.~Polosukhin, ``Attention is all you need,'' in \emph{Neural
  Information Processing Systems (NeurIPS)}, vol.~30, 2017.

\bibitem{mixit}
S.~Wisdom, E.~Tzinis, H.~Erdogan, R.~Weiss, K.~Wilson, and J.~Hershey,
  ``Unsupervised sound separation using mixture invariant training,'' in
  \emph{Neural Information Processing Systems (NeurIPS)}, vol.~33, 2020.

\bibitem{semisup_svs}
Z.~Wang, R.~Giri, U.~Isik, J.-M. Valin, and A.~Krishnaswamy, ``Semi-supervised
  singing voice separation with noisy self-training,'' \emph{IEEE International
  Conference on Acoustics, Speech and Signal Processing (ICASSP)}, 2021.

\bibitem{improved_svs}
S.~Yuan, Z.~Wang, U.~Isik, R.~Giri, J.-M. Valin, M.~Goodwin, and
  A.~Krishnaswamy, ``Improved singing voice separation with chromagram-based
  pitch-aware remixing,'' \emph{IEEE International Conference on Acoustics,
  Speech and Signal Processing (ICASSP)}, 2022.

\bibitem{remixit}
E.~Tzinis, Y.~Adi, V.~K. Ithapu, B.~Xu, and A.~Kumar, ``Continual self-training
  with bootstrapped remixing for speech enhancement,'' \emph{IEEE International
  Conference on Acoustics, Speech and Signal Processing (ICASSP)}, 2022.

\bibitem{personalized_se}
A.~Sivaraman, S.~Kim, and M.~Kim, ``Personalized speech enhancement through
  self-supervised data augmentation and purification,'' in \emph{Interspeech},
  2021.

\bibitem{sisdr}
J.~L. Roux, S.~Wisdom, H.~Erdogan, and J.~R. Hershey, ``Sdr – half-baked or
  well done?'' \emph{IEEE International Conference on Acoustics, Speech and
  Signal Processing (ICASSP)}, 2019.

\bibitem{ge2e}
L.~Wan, Q.~Wang, A.~Papir, and I.~Lopez-Moreno, ``Generalized end-to-end loss
  for speaker verification,'' \emph{IEEE International Conference on Acoustics,
  Speech and Signal Processing (ICASSP)}, 2018.

\bibitem{dprnn}
Y.~Luo, Z.~Chen, and T.~Yoshioka, ``Dual-path rnn: Efficient long sequence
  modeling for time-domain single-channel speech separation,'' \emph{IEEE
  International Conference on Acoustics, Speech and Signal Processing
  (ICASSP)}, 2020.

\bibitem{tsenet}
C.~Xu, W.~Rao, C.~E. Siong, and H.~Li, ``Time-domain speaker extraction
  network,'' \emph{IEEE Automatic Speech Recognition and Understanding Workshop
  (ASRU)}, 2019.

\bibitem{layernorm}
J.~Ba, J.~R. Kiros, and G.~E. Hinton, ``Layer normalization,'' \emph{ArXiv},
  vol. abs/1607.06450, 2016.

\bibitem{srl}
S.~Mun, S.~Choe, J.~Huh, and J.~S. Chung, ``The sound of my voice: Speaker
  representation loss for target voice separation,'' \emph{IEEE International
  Conference on Acoustics, Speech and Signal Processing (ICASSP)}, 2020.

\bibitem{voxceleb}
A.~Nagrani, J.~S. Chung, and A.~Zisserman, ``Voxceleb: A large-scale speaker
  identification dataset,'' in \emph{Interspeech}, 2017.

\bibitem{librispeech}
V.~Panayotov, G.~Chen, D.~Povey, and S.~Khudanpur, ``Librispeech: An asr corpus
  based on public domain audio books,'' in \emph{IEEE International Conference
  on Acoustics, Speech and Signal Processing (ICASSP)}, 2015.

\bibitem{adam}
D.~P. Kingma and J.~Ba, ``Adam: A method for stochastic optimization,''
  \emph{CoRR}, vol. abs/1412.6980, 2015.

\bibitem{pesq}
A.~Rix, J.~Beerends, M.~Hollier, and A.~Hekstra, ``Perceptual evaluation of
  speech quality (pesq)-a new method for speech quality assessment of telephone
  networks and codecs,'' in \emph{IEEE International Conference on Acoustics,
  Speech, and Signal Processing (ICASSP)}, vol.~2, 2001.

\end{thebibliography}
